\documentclass{article}

\usepackage{arxiv}

\usepackage[utf8]{inputenc} 
\usepackage[T1]{fontenc}    
\usepackage{hyperref}       
\usepackage{url}            
\usepackage{booktabs}       
\usepackage{amsfonts}       
\usepackage{nicefrac}       
\usepackage{microtype}      
\usepackage{lipsum}
\usepackage{graphicx}

\title{SOREL-20M: A Large Scale Benchmark Dataset for Malicious PE Detection}

\author{
  Richard Harang\thanks{Equal contribution; please direct all questions regarding the dataset to sorel-dataset@sophos.com or open an issue on GitHub for any code-related question.} \\
  Sophos AI\\
  \texttt{richard.harang@sophos.com} \\
   \And
 Ethan M. Rudd$^{*}$\thanks{Ethan Rudd is currently with FireEye Inc. All RTD\&E associated with this effort conducted by this author was performed while working for Sophos AI.}\\
    FireEye Data Science\\
  \texttt{ethan.rudd@fireeye.com} \\
}

\begin{document}
\maketitle

\begin{abstract}
In this paper we describe the SOREL-20M (Sophos/ReversingLabs-20 Million) dataset: a large-scale dataset consisting of nearly 20 million files with pre-extracted features and metadata, high-quality labels derived from multiple sources, information about vendor detections of the malware samples at the time of collection, and additional ``tags'' related to each malware sample to serve as additional targets.  In addition to features and metadata, we also provide approximately 10 million ``disarmed'' malware samples -- samples with both the optional\_headers.subsystem and file\_header.machine flags set to zero -- that may be used for further exploration of features and detection strategies.  We also provide Python code to interact with the data and features, as well as baseline neural network and gradient boosted decision tree models and their results, with full training and evaluation code, to serve as a starting point for further experimentation. 
\end{abstract}

\section{Introduction: Why Another Dataset?}
The use of machine learning for malware detection is now relatively widespread.  The ability for modern machine learning models to learn complex relationships between a large number of both statistical and parse-based features has lead to their widespread adoption.  However, the risks associated with working directly with malware, as well as the commercial nature of much research in this space, has meant that most ML-based malware models are evaluated on private or proprietary datasets. This makes measuring progress in the field difficult. Furthermore, many sources of malware are commercial in nature, placing a high barrier of entry to the field and leading to researchers evaluating their models on extremely small datasets.

In contrast, fields like image classification or natural language processing have arguably benefited immensely from large, publicly available datasets such as CIFAR \cite{krizhevsky2009learning}, ImageNet \cite{deng2009imagenet}, or the Stanford Sentiment Treebank \cite{socher2013recursive}, which allow researchers to apply different approaches on a common dataset, making a direct comparison of those approaches possible.  In addition to providing a basis for comparison between approaches, the existence of these common datasets has also made the fields more accessible, allowing smaller organizations that lacked the ability to compile large training and validation sets to contribute to the development of those fields.  

The first attempt to address this lack was the seminal EMBER dataset \cite{anderson2018ember}, which the present work builds upon.  The EMBER dataset was the first standard dataset to be used for malware detection, however it had some shortcomings that limited its utility as a malware benchmark set.  First, EMBER was of limited size, containing 900,000 training samples and 200,000 test samples, while commercial malware models are trained on tens to hundreds of millions of samples.  In addition to the training size being too small to compare to commercial scale, the small validation size makes evaluation of model performance at lower false positive rates (1 in 1000 or below) difficult due to variance issues. Perhaps due to the relatively small size of the dataset, performance of classifiers on EMBER is nearly saturated, with a baseline classifier capable of obtaining an AUC of over 0.999 \cite{anderson2018ember}. In addition, the EMBER dataset provided only pre-extracted features, making further research in such topic as improvements in feature extraction or realizable adversarial sample generation difficult.  Finally, EMBER provides only a single binary label based on a simple `thresholding' rule.  

In the hopes of being a valuable benchmark set for malware detection, SOREL-20M attempts to address these issues in whole or in part.  We address the issue of training size by providing an order of magnitude more samples for analysis.  Internally, we have found that while performance continues to improve with larger datasets, validation sizes on the order of 3 to 4 million examples are sufficient to establish a stable rank order between models as well as to assess performance at lower false positive rates.  If our recommended time splits \cite{berlin2016improving} are used to establish training, validation, and test sets, we obtain 12,699,013 training samples, 2,495,822 validation samples, and 4,195,042 test samples, respectively.  This is sufficient to ensure that comparisons of different models, architectures, and features can establish relative performance; particularly if care is taken to examine model variance at the same time using multiple random initializations of the model.

We partially address the issue of feature exploration by providing (disarmed) binary samples for the malware only. In all, we provide 9,919,251 binary samples of malware (7,596,407 training samples, 962,222 validation samples, and 1,360,622 test samples), which have been `disarmed' by setting both the optional\_headers.subsystem and file\_header.machine flags to 0 in order to prevent execution.  We have also provided complete PE metadata as obtained via the Python \textit{pefile} \cite{carrera2007win32} module using the dump\_dict() method.  While this does hinder direct comparisons of models, comparisons of the distribution of scores on the malware set using researcher-provided feature extraction code does still allow for comparison of detection rates at different thresholds.

Simiarly to EMBER, we have established baseline models on SOREL-20M using both LightGBM \cite{ke2017lightgbm} and a PyTorch \cite{paszke2019pytorch} based feed-forward neural network (FFNN) model. While performance is high for both models, there remains significant room for improvement, particularly at the lower false positive rates (which the large size of the SOREL-20M corpus now allows us to evaluate with confidence).  We anticipate that this will make SOREL-20M more useful as a method of comparing malware detection approaches to each other.  Finally, we provide a number of additional targets for the model that describe behaviors inferred from vendor labels (as described in \cite{ducau2019smart}) which we also provide benchmarks for using a multi-target model as described in \cite{rudd2019aloha}.

In the following sections, we describe the corpus statistics and structure of the data and how it may be accessed.  We then describe the baseline models we have trained on the data, as well as the layout of the associated GitHub repository.

\section{Dataset description}
\label{sec:corpus_description}

The complete dataset consists of the following items:
\begin{itemize}
  \item The 9,919,251 original (disarmed) malware samples; available via S3 at s3://sorel-20m/09-DEC-2020/binaries/ compressed via the python zlib.compress function
  \item A SQLite3 and two LMDB databases, available via S3 at s3://sorel-20m/09-DEC-2020/processed-data/ 
    \begin{itemize}
        \item The SQLite3 ``meta.db'' database contains malware labels, tags, detection counts, and first/last seen times
        \item The ``ember\_features'' LMDB database contains EMBER features (extracted with version 2 of the features)
        \item the ``pe\_metadata'' LMDB database contains the PE metadata extracted via the pefile module, as described above
    \end{itemize}  
    \item Pre-trained baseline models and results, available via S3 at s3://sorel-20m/09-DEC-2020/baselines/ 
\end{itemize}

We also provide Python code at https://github.com/sophos-ai/SOREL-20M (see section \ref{sec:github}) that will interact with the SQLite and LMDB combination databases that are provided, and can be used to train the baseline models that we provide.

All samples are identified by sha256; in the case of the disarmed malware samples, we use the sha256 of the original, unmodified file, and not the sha256 of the disarmed file.  The sha256 serves as the primary key for the SQLite database, and the key for the two LMDB databases.  LMDB entries are stored as arrays or dictionaries (for EMBER feature vectors or PE metadata, respectively) that are then serialized with msgpack and compressed with zlib.  

The data was collected from January 1, 2017 to April 10, 2019.  We suggest time-splits of the data -- based upon the first-seen time in RL telemetry -- as follows: training data from the beginning of collection until November 29, 2018; validation data from then until January 12, 2019; and testing data from January 12, 2019 through the end of the data.  Using those time splits, the breakdown of malicious and benign samples in the training, validation, and testing sets is given in table \ref{tbl:corpus_stats}.

\begin{table}[]
\centering
\label{tbl:corpus_stats}
\begin{tabular}{l|r|r|}
\cline{2-3}
& \multicolumn{1}{c|}{\textbf{Malicious}} & \multicolumn{1}{c|}{\textbf{Benign}} \\ \hline
\multicolumn{1}{|l|}{\textbf{Training set}}   & 7596407                                 & 5102606                              \\ \hline
\multicolumn{1}{|l|}{\textbf{Validation set}} & 962222                                  & 1533579                              \\ \hline
\multicolumn{1}{|l|}{\textbf{Test set}}       & 1360622                                 & 2834441                              \\ \hline
\end{tabular}
\caption{Distribution of malware and benign samples across our suggested training, testing, and validation splits}
\end{table}

\begin{figure}
  \centering
    \includegraphics[width=0.4\linewidth]{"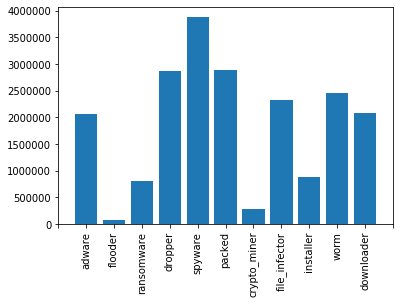"}
  \caption{The distribution of behavioral tags in the training set.}
  \label{fig:train_tags}
\end{figure}

\begin{figure}
  \centering
    \includegraphics[width=0.4\linewidth]{"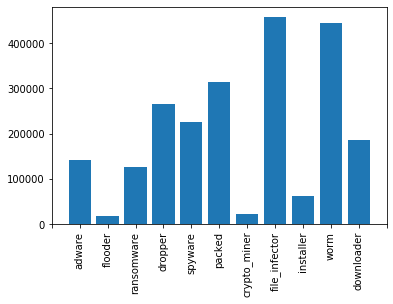"}
  \caption{The distribution of behavioral tags in the recommended validation set.}
  \label{fig:validation_tags}
\end{figure}

\begin{figure}
  \centering
    \includegraphics[width=0.4\linewidth]{"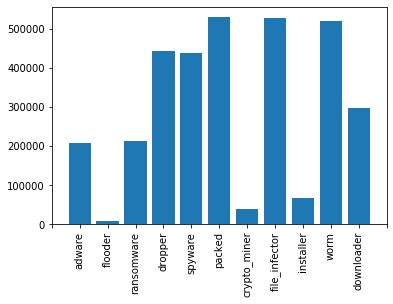"}
  \caption{The distribution of behavioral tags in the recommended test set.}
  \label{fig:test_tags}
\end{figure}

The corpus statistics for the behavioral tags are give in figures \ref{fig:train_tags}, \ref{fig:validation_tags}, and \ref{fig:test_tags}.


\section{Baseline Models}
\label{sec:models}
We provide two baseline models; a Pytorch feed-forward neural network (FFNN) model, and a LightGBM gradient-boosted decision tree model.  Both models are trained on the EMBER-v2 features available in the LMDB described in section \ref{sec:corpus_description}, and trained using code from the GitHub respository described in section \ref{sec:github} using random seeds of 1, 2, 3, 4, and 5, respectively.

The FFNN model is a simplified version of the model from \cite{rudd2019aloha} consisting of three `blocks' followed by one or more output `heads'.  A block consists of a Linear layer, LayerNorm, ELU activation, and Dropout.  The output heads consist of additional linear and activation layers to produce outputs for tags, counts, and malware classification.  See the file `nets.py' in the github repository for full details.  

The LightGBM model is trained with 500 iterations, unbounded maximum depth but a maximum of 64 leaves, with a bagging fraction of 0.9,  a feature subselection both at a tree level and at a node level of 0.9, and an early stopping rounds count of 10 (see `lightgbm\_config.json' in the GitHub repositiory for all parameters).  The helper script `build\_numpy\_arrays\_for\_lightgbm.py' may be used to unload our LMDB features into numpy arrays suitable for training the LightGBM model, however this requires both significant disk space to hold the files (approximately 175GB in total) and a similar amount of RAM during training of the LightGBM model; we used an AWS  m5.24xlarge instance for both tasks.

\begin{figure}
  \centering
    \includegraphics[width=0.8\linewidth]{"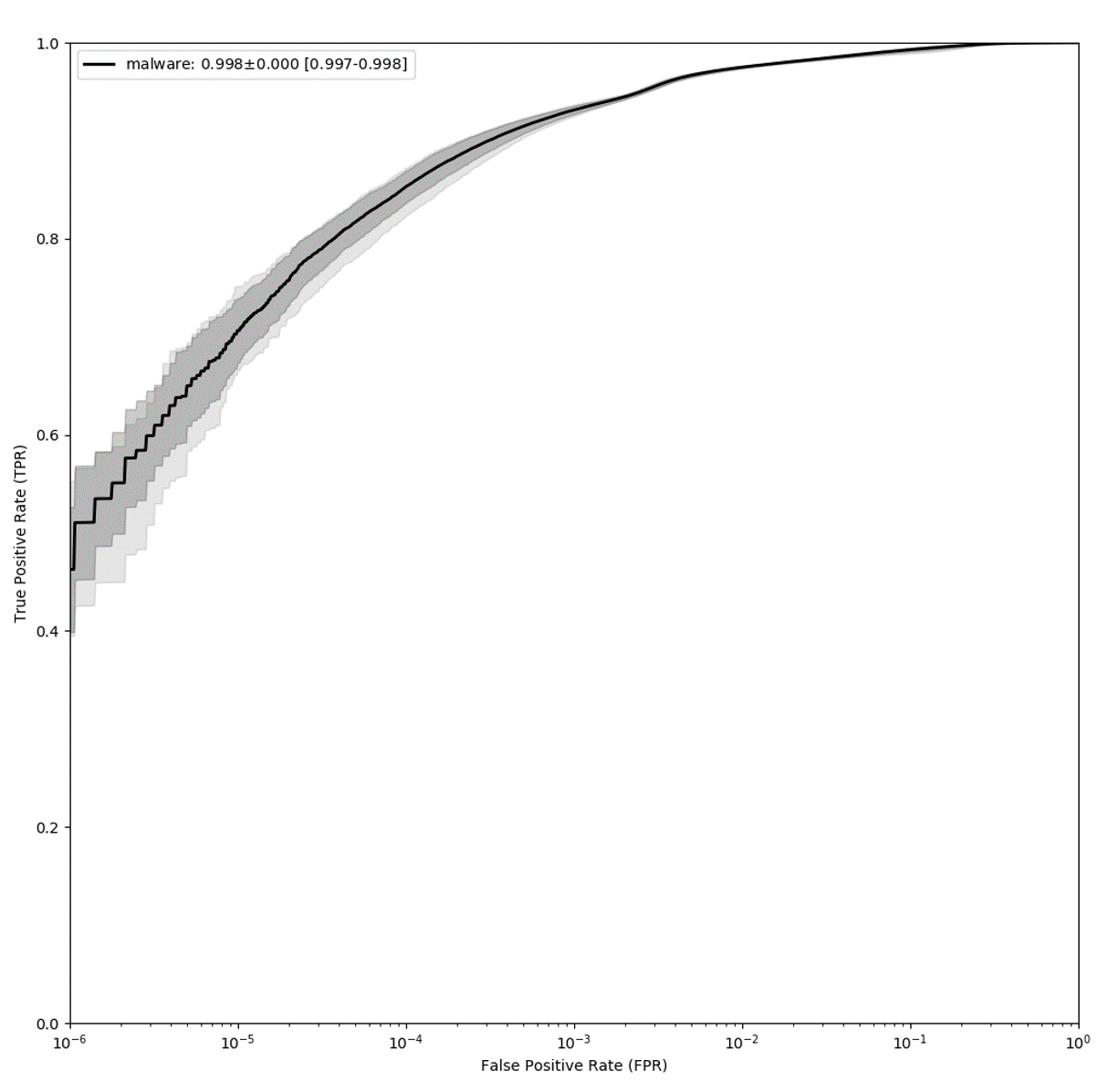"}
  \caption{ROC for FFNN with statistics aggregated over five trials; mean shown as black line; dark region indicates plus/minus one standard deviation; light region indicates min/max.}
  \label{fig:ffnn_roc}
\end{figure}

\begin{figure}
  \centering
    \includegraphics[width=0.8\linewidth]{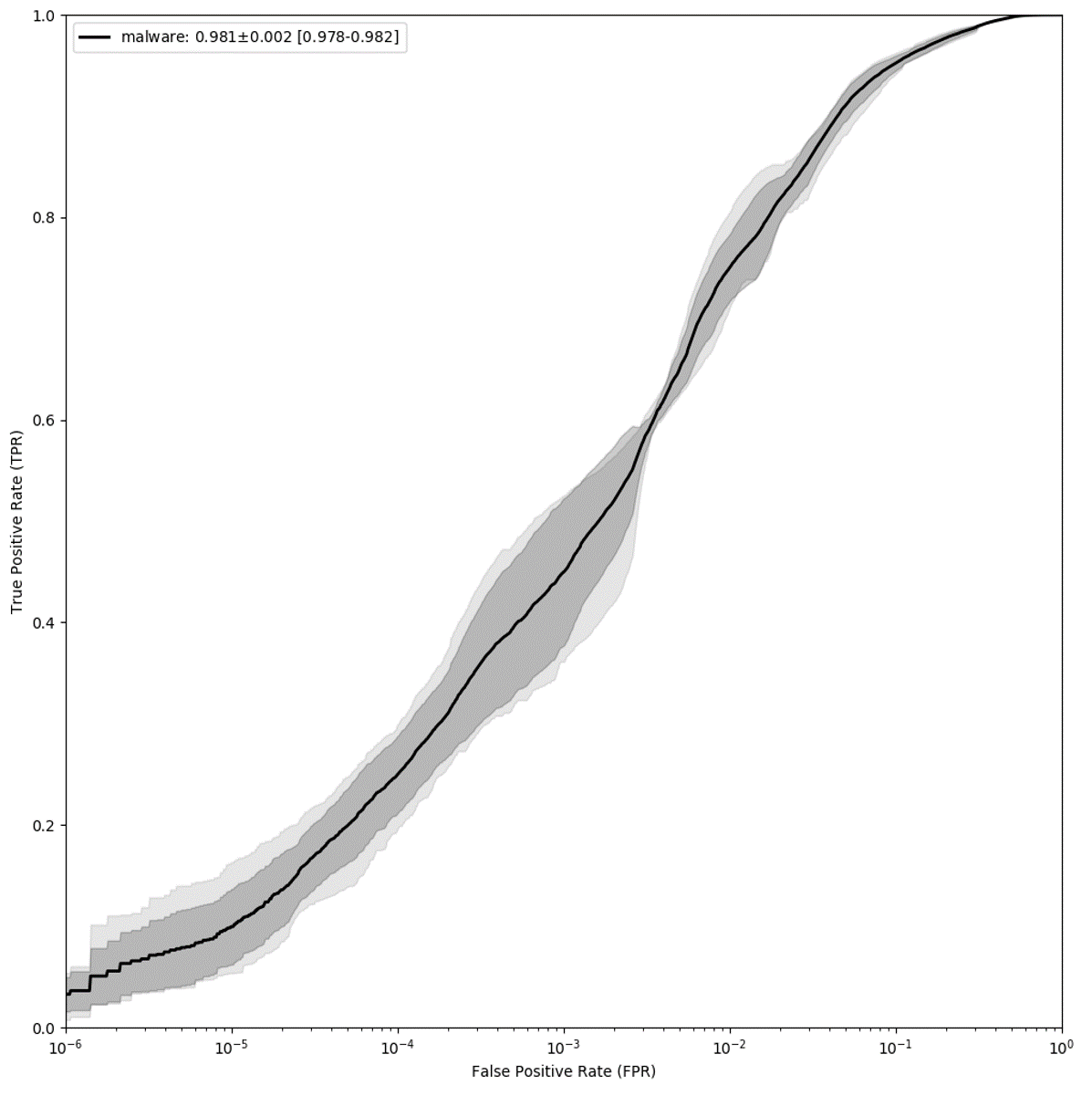}
  \caption{ROC for LightGBM GBDT with statistics aggregated over five trials; mean shown as black line; dark region indicates plus/minus one standard deviation; light region indicates min/max.}
  \label{fig:lightgbm_roc}
\end{figure}

\begin{figure}
    \centering
    \includegraphics[width=0.8\linewidth]{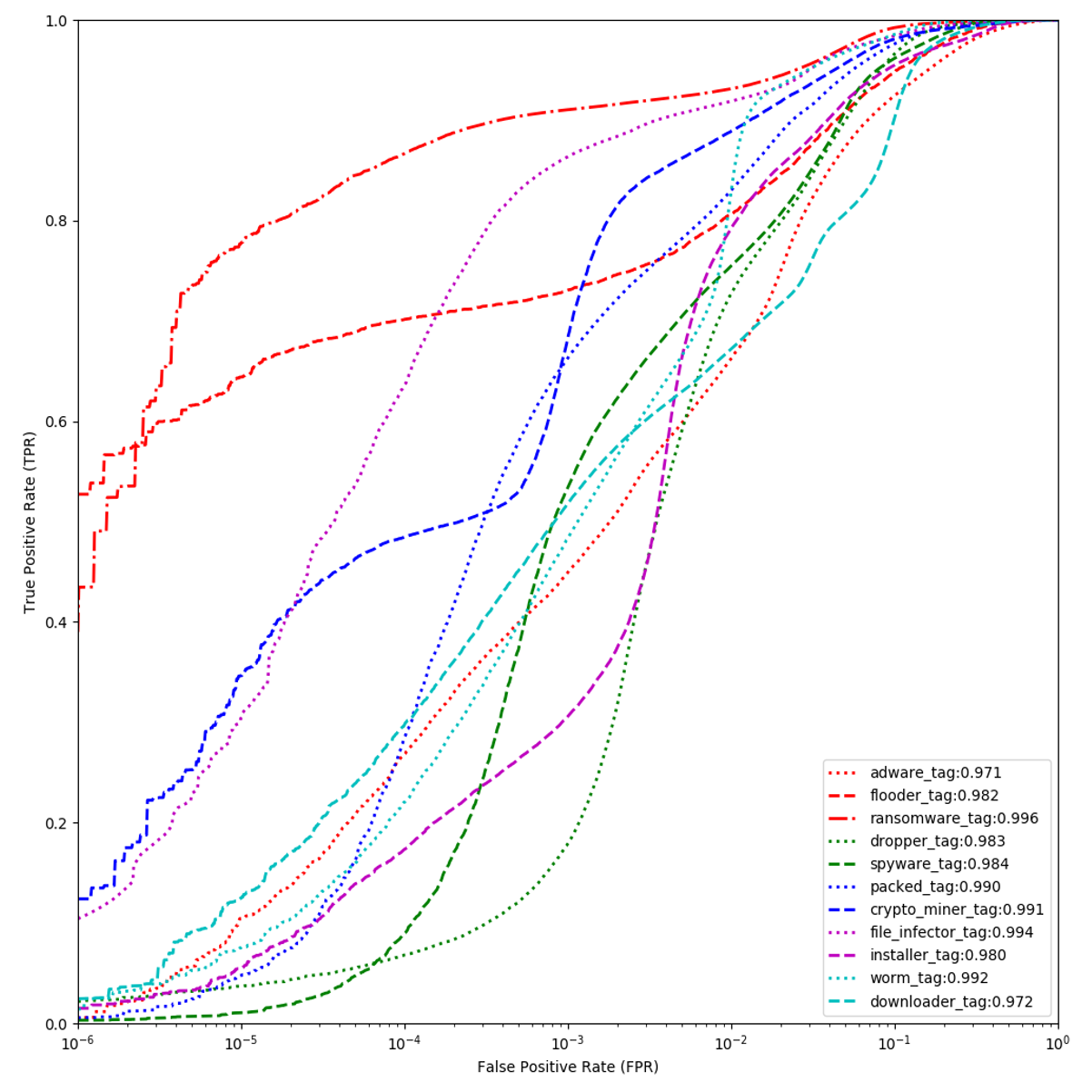}
    \caption{ROC for various tag predictions using the FFNN model with predict\_tags set to `True'}
    \label{fig:ffnn_tags}
\end{figure}

ROC plots for the models are given in figures \ref{fig:ffnn_roc} (FFNN ROC for malware output), \ref{fig:lightgbm_roc} (LightGBM ROC), and \ref{fig:ffnn_tags} (individual per-tag ROCs for the tags in the FFNN model).  Note that the ROC for the malware output and tags in the FFNN model are obtained from the same mode trained using the tags in a multi-target learning setting \cite{caruana1998dozen}, which we have observed to improve the overall performance of the malware output (see \cite{rudd2019aloha}).  As multi-target learning is not implemented in LightGBM it is trained on the single malware task, which may be in part why the performance is lower.


\section{GitHub Repository}
\label{sec:github}
The GitHub repository containing supporting code may be accessed at https://github.com/sophos-ai/SoReL-20M and is licensed under Apache 2.0 license.  It contains the code which was used to train the provided baseline models, code which may be used to interact with the databases containing pre-extracted features and metadata, and supporting files including an Anaconda \cite{anaconda} compatible YaML file describing a minimal environment for building the models and a list of sha256 values for which EMBER-v2 features could not be extracted to accelerate dataset load times.

Complete details on initializing the environment and training the models are provided in the README file in the repository, via the `--help' option in the various scripts, and function docstrings.  Briefly; the `train.py' and `evaluate.py' files train and evaluate the models, respectively.  The `plot.py' command takes in a JSON file denoting which runs to plot (an example is provided in the S3 bucket with the pretrained model weights) and outputs plots of the form shown in section \ref{sec:models}.  The pytorch FFNN model is specified in `nets.py' and the LightGBM model is specified in the file `lightgbm\_config.json'.  

The `dataset.py' and `generators.py' functions are of most interest to those who wish to use other frameworks to train a model.  The Dataset class in `dataset.py' subclasses the PyTorch Dataset class to link the SQLite3 meta.db file to the feature LMDBs.  The GeneratorFactory class in `generators.py' provides an interface to load a dataset.Dataset instance and wrap it in a PyTorch DataLoader in preparation for training.  Finally, as noted in section \ref{sec:models}, the file `build\_numpy\_arrays\_for\_lightgbm.py' is a command-line utility to iterate over a generator and write it to a Numpy \cite{numpy} .npz file which may then be used to train a LightGBM model.

\section{Conclusion}
We have presented SOREL-20M dataset, which includes: 
\begin{itemize}
    \item nearly 10 million disarmed but otherwise complete malware files\footnote{Benign files are excluded due to potential intellectual property concerns.}
    \item extracted features and metadata for 20 million malicious and benign portable executable files
    \item extensive metadata including behavior-like tags, number of detections,  and high-quality internally developed and validated malware/benignware labels for all 20 million files
    \item a set of 10 pre-trained models to serve as a baseline
    \item complete source code required to reproduce our results and explore further developments using the data
\end{itemize}

To our knowledge this is both the largest malware benchmark training set that has been released to date, as well as the first to contain a reference set of malware observed ``in the wild'' comparatively recently.  This dataset allows for ``fair'' comparisons between different models using sufficient data to allow good comparisons at relevant false positive rates, as well as evaluation of performance scores on a reference set of malware using novel and arbitrary researcher-developed features.  

Sophos and ReversingLabs are proud to offer this dataset in hopes it will further stimulate the development of the field.

\bibliographystyle{unsrt}  


\begin{thebibliography}{1}
 

\bibitem{krizhevsky2009learning}
Alex Krizhevsky, Geoffrey Hinton, et~al.
\newblock Learning multiple layers of features from tiny images.
\newblock 2009.

\bibitem{deng2009imagenet}
Jia Deng, Wei Dong, Richard Socher, Li-Jia Li, Kai Li, and Li~Fei-Fei.
\newblock Imagenet: A large-scale hierarchical image database.
\newblock In {\em 2009 IEEE conference on computer vision and pattern
  recognition}, pages 248--255. Ieee, 2009.

\bibitem{socher2013recursive}
Richard Socher, Alex Perelygin, Jean Wu, Jason Chuang, Christopher~D Manning,
  Andrew~Y Ng, and Christopher Potts.
\newblock Recursive deep models for semantic compositionality over a sentiment
  treebank.
\newblock In {\em Proceedings of the 2013 conference on empirical methods in
  natural language processing}, pages 1631--1642, 2013.

\bibitem{anderson2018ember}
Hyrum~S Anderson and Phil Roth.
\newblock Ember: an open dataset for training static pe malware machine
  learning models.
\newblock {\em arXiv preprint arXiv:1804.04637}, 2018.

\bibitem{berlin2016improving}
Konstantin Berlin and Joshua Saxe.
\newblock Improving zero-day malware testing methodology using statistically
  significant time-lagged test samples.
\newblock {\em arXiv preprint arXiv:1608.00669}, 2016.

\bibitem{carrera2007win32}
Ero Carrera.
\newblock Win32 static analysis in python, 2007.

\bibitem{ke2017lightgbm}
Guolin Ke, Qi~Meng, Thomas Finley, Taifeng Wang, Wei Chen, Weidong Ma, Qiwei
  Ye, and Tie-Yan Liu.
\newblock Lightgbm: A highly efficient gradient boosting decision tree.
\newblock In {\em Advances in neural information processing systems}, pages
  3146--3154, 2017.

\bibitem{paszke2019pytorch}
Adam Paszke, Sam Gross, Francisco Massa, Adam Lerer, James Bradbury, Gregory
  Chanan, Trevor Killeen, Zeming Lin, Natalia Gimelshein, Luca Antiga, et~al.
\newblock Pytorch: An imperative style, high-performance deep learning library.
\newblock In {\em Advances in neural information processing systems}, pages
  8026--8037, 2019.

\bibitem{ducau2019smart}
Felipe~N Ducau, Ethan~M Rudd, Tad~M Heppner, Alex Long, and Konstantin Berlin.
\newblock Smart: Semantic malware attribute relevance tagging.
\newblock {\em CoRR}, 2019.

\bibitem{rudd2019aloha}
Ethan~M Rudd, Felipe~N Ducau, Cody Wild, Konstantin Berlin, and Richard Harang.
\newblock Aloha: Auxiliary loss optimization for hypothesis augmentation.
\newblock In {\em 28th $\{$USENIX$\}$ Security Symposium ($\{$USENIX$\}$
  Security 19)}, pages 303--320, 2019.

\bibitem{caruana1998dozen}
Rich Caruana.
\newblock A dozen tricks with multitask learning.
\newblock In {\em Neural networks: tricks of the trade}, pages 165--191.
  Springer, 1998.

\bibitem{anaconda}
Anaconda software distribution, 2020.

\bibitem{numpy}
Travis Oliphant.
\newblock {NumPy}: A guide to {NumPy}.
\newblock USA: Trelgol Publishing, 2006--.
\newblock [Online; accessed <today>].

\end{thebibliography}


\end{document}